\documentstyle[onecolumn,epsf]{mn}

\newif\ifAMStwofonts

\ifoldfss
  \ifCUPmtlplainloaded \else
    \NewTextAlphabet{textbfit} {cmbxti10} {}
    \NewTextAlphabet{textbfss} {cmssbx10} {}
    \NewMathAlphabet{mathbfit} {cmbxti10} {} 
    \NewMathAlphabet{mathbfss} {cmssbx10} {} 
  \fi
  \ifAMStwofonts
    \ifCUPmtlplainloaded \else
      \NewSymbolFont{upmath} {eurm10}
      \NewSymbolFont{AMSa} {msam10}
      \NewMathSymbol{\upi}     {0}{upmath}{19}
      \NewMathSymbol{\umu}     {0}{upmath}{16}
      \NewMathSymbol{\upartial}{0}{upmath}{40}
      \NewMathSymbol{\leqslant}{3}{AMSa}{36}
      \NewMathSymbol{\geqslant}{3}{AMSa}{3E}

    \fi
  \fi
\fi 

\ifnfssone
  \newmathalphabet{\mathit}
  \addtoversion{normal}{\mathit}{cmr}{m}{it}
  \addtoversion{bold}{\mathit}{cmr}{bx}{it}
  \newmathalphabet{\mathbfit} 
  \addtoversion{normal}{\mathbfit}{cmr}{bx}{it}
  \addtoversion{bold}{\mathbfit}{cmr}{bx}{it}
  \newmathalphabet{\mathbfss} 
  \addtoversion{normal}{\mathbfss}{cmss}{bx}{n}
  \addtoversion{bold}{\mathbfss}{cmss}{bx}{n}
  \ifAMStwofonts
    \ifCUPmtlplainloaded \else
      \UseAMStwoboldmath
      \makeatletter
      \new@mathgroup\upmath@group
      \define@mathgroup\mv@normal\upmath@group{eur}{m}{n}
      \define@mathgroup\mv@bold\upmath@group{eur}{b}{n}
      \edef\UPM{\hexnumber\upmath@group}
      \new@mathgroup\amsa@group
      \define@mathgroup\mv@normal\amsa@group{msa}{m}{n}
      \define@mathgroup\mv@bold\amsa@group{msa}{m}{n}
      \edef\AMSa{\hexnumber\amsa@group}
      \makeatother
      \mathchardef\upi="0\UPM19
      \mathchardef\umu="0\UPM16
      \mathchardef\upartial="0\UPM40
      \mathchardef\leqslant="3\AMSa36
      \mathchardef\geqslant="3\AMSa3E
    \fi
  \fi
\fi 

\ifnfsstwo
  \DeclareMathAlphabet{\mathbfit}{OT1}{cmr}{bx}{it}
  \SetMathAlphabet\mathbfit{bold}{OT1}{cmr}{bx}{it}
  \DeclareMathAlphabet{\mathbfss}{OT1}{cmss}{bx}{n}
  \SetMathAlphabet\mathbfss{bold}{OT1}{cmss}{bx}{n}
  \ifAMStwofonts
    \ifCUPmtlplainloaded \else
      \DeclareSymbolFont{UPM}{U}{eur}{m}{n}
      \SetSymbolFont{UPM}{bold}{U}{eur}{b}{n}
      \DeclareSymbolFont{AMSa}{U}{msa}{m}{n}
      \DeclareMathSymbol{\upi}{0}{UPM}{"19}
      \DeclareMathSymbol{\umu}{0}{UPM}{"16}
      \DeclareMathSymbol{\upartial}{0}{UPM}{"40}
      \DeclareMathSymbol{\leqslant}{3}{AMSa}{"36}
      \DeclareMathSymbol{\geqslant}{3}{AMSa}{"3E}
    \fi
  \fi
\fi 

\ifCUPmtlplainloaded \else
  \ifAMStwofonts \else 
    \def\upi{\pi}
    \def\umu{\mu}
    \def\upartial{\partial}
  \fi
\fi

\title{Extending Lagrangian perturbation theory
to a fluid with velocity dispersion}

\author[M. Morita and T. Tatekawa]
       {Masaaki Morita$^{1,2}$ and Takayuki Tatekawa$^3$ \\
        $^1$Department of Physics, Ochanomizu University,
            Ohtsuka, Bunkyo, Tokyo 112-8610, Japan\\
        $^2$Advanced Research Institute for Science and Engineering,
            Waseda University, Ohkubo, Shinjuku, Tokyo 169-8555, Japan\\
        $^3$Department of Physics, Waseda University,
            Ohkubo, Shinjuku, Tokyo 169-8555, Japan}

\date{2001 July 4}

\pagerange{\pageref{firstpage}--\pageref{lastpage}}

\pubyear{2001}

\begin{document}
\maketitle
\label{firstpage}

\begin{abstract}
We formulate a perturbative approximation to gravitational instability,
based on Lagrangian hydrodynamics in Newtonian cosmology.
We take account of `pressure' effect of fluid,
which is kinematically caused by velocity dispersion,
to aim hydrodynamical description beyond shell crossing.
Master equations in the Lagrangian description are derived
and solved perturbatively up to second order.
Then, as an illustration,
power spectra of density fluctuations are computed
in a one-dimensional model from the Lagrangian approximations
and Eulerian linear perturbation theory for comparison.
We find that the results by the Lagrangian approximations
are different from those by the Eulerian one in weakly non-linear regime
at the scales smaller than the Jeans length.
We also show the validity of the perturbative Lagrangian approximations
by consulting difference between the first-order
and the second-order approximations.
\end{abstract}


\section{Introduction}

It is significant to investigate evolution of inhomogeneities
by gravitational instability in the expanding universe from
the viewpoint of cosmological structure formation.
In order to find how to form cosmic structures
via gravitational instability,
numerical simulations such as $N$-body simulations
have been carried out by several groups
(e.g. Miyoshi \& Kihara 1975; Hockney \& Eastwood 1988; Couchman 1999).
Such numerical approaches have brought us many useful informations
about structure formation,
but analytical approaches are also needed to obtain physical understanding
of structure formation.

 For analytical approaches, one usually treats matter
contained in the universe as a self-gravitating fluid,
and considers solving the hydrodynamical equations for the fluid.
Since the hydrodynamical equations are generally non-linear,
one cannot solve them without any assumption or approximation.
A conventional approximation is linear perturbation
in homogeneous and isotropic universes,
based on the Eulerian picture of hydrodynamics
\cite{weinberg,peebles,coles,saco}.
This approach is, by construction, valid only in linear regime,
where amplitude of density perturbations is much smaller than unity.
 For description beyond linear regime,
Zel'dovich \shortcite{zel} proposed a new approximation scheme
in which perturbations are given as the Lagrangian displacement
of fluid flow by an extrapolation of the linear perturbation theory.
This approximation is known as Zel'dovich approximation,
which has been found to give relatively accurate results
and work better than the Eulerian approximations
by comparison with exact solutions \cite{munshi,sasha,yoshisato},
in weakly non-linear regime,
where amplitude of density perturbations becomes comparable to unity.
The Zel'dovich approximation is shown to be a subclass
of the first-order solutions of a perturbation theory
in the Lagrangian hydrodynamics (Buchert 1989, 1992),
and along this line, higher-order extensions have been developed
up to third order
\cite{bouchet92,bueh93,buchert94,bouchet95,catelan,sasakasa}.

In the Zel'dovich approximation, however,
physical singularities called `shell crossing' inevitably occur.
This is a consequence due to the fact
that a self-gravitating pressureless fluid is taken
as a matter model in the approximation.
At the epoch beyond shell crossing,
the Zel'dovich approximation soon becomes inaccurate
because the fluid elements move throughout in the directions
which are set initially,
and then inhomogeneous structures, which are formed compactly once,
are dissolved in the approximation scheme.
To resolve this problem, some modifications have been proposed,
such as `truncated Zel'dovich approximation' \cite{cms,mps},
which is an optimization of the approximation
by truncating small-scale fluctuations,
and `adhesion approximation' \cite{gurbatov},
where an artificial viscosity is introduced.
These modifications eliminate the shortcomings
of the Zel'dovich approximation,
and actually the modified approximations provide excellent results
compared with $N$-body simulations in some cases.
However the physical grounds of the modifications are not clarified.

To have more well-founded approximations from a physical point of view,
we need to study gravitational instability of pressureless matter
beyond shell crossing.
Buchert \& Dom\'{\i}nguez \shortcite{budo} have examined this issue,
starting from the collisionless Boltzmann equation,
which is usually applied to the stellar systems
(e.g. Binney \& Tremaine 1987).
They argued that effect of velocity dispersion will be significant
beyond shell crossing,
and if the velocity dispersion is approximately isotropic,
it yields pressure-like or viscosity terms.
This implies that the gravitational instability of pressureless matter
beyond shell crossing can be described effectively
by hydrodynamic equations for a fluid with pressure-like force.
 Following this view, Adler \& Buchert \shortcite{adler} have
proposed reformulation of the Lagrangian perturbation theory
by taking account of pressure effect.
They derived first-order perturbation equations
in the Lagrangian coordinates
under the assumption that the pressure is a function of only mass density.
They did not, however, present solutions of the perturbation equations
or analyse the evolution of density perturbations with the solutions.
One may expect the reformulation to extend the regions
which can be described by analytical approximations,
but this should be confirmed by a concrete illustration.
The aim of this paper is to show how the reformulation gives
description of gravitational instability
by solving perturbation equations and illustrating behaviour
of density perturbations.

In this paper, we derive and solve the Lagrangian perturbation equations
with pressure up to second order, assuming a polytropic equation of state.
We adopt the method of the Fourier transformation for the solutions
and then will see mode couplings in the Lagrangian Fourier space
in the second-order solutions.
In particular, we obtain explicit form of the second-order solutions
in the case $P \propto \rho^{4/3}$,
where $P$ and $\rho$ are pressure and mass density, respectively.
Moreover, as an illustration of the formulation,
power spectra of density fluctuations are computed
in a one-dimensional model from the Lagrangian approximations
for the case $P \propto \rho^{4/3}$,
and are compared with the results by the Eulerian linear
perturbation theory to clarify the difference between them.
We also compare the first-order and the second-order Lagrangian
approximations to examine the validity of the approximation scheme.

This paper is organized as follows.
In Section 2 we present basic equations of our method.
Starting from the hydrodynamical equations,
we derive master equations of the Lagrangian perturbation theory
with pressure effect.
In Section 3 we obtain perturbation equations
by expanding the master equations up to second order,
and solve them via the Fourier transformation.
Section 4 gives illustrative examples of computation
of density perturbations
by the Lagrangian and the Eulerian approximations
in a one-dimensional model.
Showing power spectra of density perturbations,
we discuss differences among the approximations.
Section 5 contains concluding remarks.

\section{Basic equations}

We begin with basic equations of cosmological hydrodynamics
for a self-gravitating fluid with energy density $\rho$
and `pressure' $P$.
In coordinates $\bmath{x} \equiv \bmath{r}/a$
comoving with cosmic expansion, they are
\begin{equation}\label{contin}
\frac{\partial \rho}{\partial t} + 3\frac{\dot a}{a}\rho
+ \frac{1}{a}\nabla_{\bmath{x}} \cdot (\rho \bmath{v}) =0 \,,
\end{equation}
\begin{equation}\label{euler}
\frac{\partial \bmath{v}}{\partial t} + \frac{\dot a}{a}\bmath{v}
+ \frac{1}{a}(\bmath{v} \cdot \nabla_{\bmath{x}})\bmath{v}
= \bmath{g} - \frac{1}{\rho a}\nabla_{\bmath{x}} P \,,
\end{equation}
\begin{equation}\label{curlfree}
\nabla_{\bmath{x}} \times \bmath{g} = 0 \,,
\end{equation}
\begin{equation}\label{poisson}
\nabla_{\bmath{x}} \cdot \bmath{g}
= -4\pi Ga (\rho - \rho_{\rm b}) \,,
\end{equation}
where $a=a(t)$ is the cosmic scale factor,
$\rho_{\rm b}=\rho_{\rm b}(t)$ is energy density of
a homogeneous and isotropic (background) universe,
and $\bmath{v}$ and $\bmath{g}$ represent respectively
velocity field and gravitational field strength
due to presence of inhomogeneity,
and then may be called as `peculiar velocity field'
and `peculiar gravitational field.'
The `pressure' we take into account here is kinematical one
due to occurrence of velocity dispersion beyond shell crossing
of dust flow,
as stated by Buchert \& Dom\'{\i}nguez \shortcite{budo},
rather than thermodynamical one.
Thus equation~(\ref{euler}) is close to the Jeans equation,
which is gained by taking moments of the collisionless
Boltzmann equation (e.g. Binney \& Tremaine 1987).

In Lagrangian description of hydrodynamics,
using the time derivative along the fluid flow
\[
\frac{{\rm d}}{{\rm d} t} \equiv
\frac{\partial}{\partial t}
+ \frac{1}{a}(\bmath{v} \cdot \nabla_{\bmath{x}}) \,,
\]
equations~(\ref{contin}) and (\ref{euler}) are rewritten as
\begin{equation}\label{dotrhoL}
\frac{{\rm d} \rho}{{\rm d} t} + 3\frac{\dot a}{a}\rho 
+ \frac{\rho}{a}(\nabla_{\bmath{x}} \cdot \bmath{v}) = 0 \,,
\end{equation}
\begin{equation}\label{dotvL}
\frac{{\rm d} \bmath{v}}{{\rm d} t} + \frac{\dot a}{a}\bmath{v}
= \bmath{g} - \frac{1}{\rho a}\nabla_{\bmath{x}} P \,.
\end{equation}
The coordinates $\bmath{x}$ of trajectories of the fluid elements
are expressed by Lagrangian coordinates $\bmath{q}$,
defined by initial values of the coordinates $\bmath{x}$, in the form
\begin{equation}\label{x=q+s}
\bmath{x} = \bmath{q} + \bmath{s} (\bmath{q},t) \,,
\end{equation}
where $\bmath{q}$ and $\bmath{s}$ represent
the background Hubble flow and
deviation of the flow from the background, respectively.
The continuity equation~(\ref{dotrhoL}) is then exactly solved as
\begin{equation}\label{exactrho}
\rho = \rho_{\rm b} J^{-1} \,,
\end{equation}
or equivalently for density contrast
$\delta \equiv (\rho - \rho_{\rm b})/\rho_{\rm b}$,
\begin{equation}\label{exactdelta}
\delta = J^{-1} -1 \,,
\end{equation}
where
$J \equiv \det (\partial x_i/\partial q_j)
        = \det (\delta_{ij} + \partial s_i/\partial q_j)$
is the Jacobian of the transformation $\bmath{x} \rightarrow \bmath{q}$.
The peculiar velocity is written by definition as
$\bmath{v} = a\dot{\bmath{s}}$ \,,
and from equation~(\ref{euler}) the peculiar gravitational field becomes
\[
\bmath{g} = a \left(\ddot{\bmath{s}} + 2\frac{\dot a}{a}\dot{\bmath{s}}
- \frac{1}{a^2} \frac{{\rm d} P}{{\rm d} \rho}
J^{-1} \nabla_{\bmath{x}} J \right)\,,
\]
where an overdot $(\dot{\ })$ denotes ${\rm d}/{\rm d} t$.
Note that the square of the `sound speed,' ${\rm d} P/{\rm d} \rho$, is
a function of $\rho$, and can be written in terms of $\bmath{s}$
by using equation~(\ref{exactrho}) if an equation of state is provided.
Now all physical quantities are found to be written in terms of $\bmath{s}$
and it remains only to find solutions for $\bmath{s}$.
We obtain the following master equations for $\bmath{s}$ from
equations~(\ref{curlfree}) and (\ref{poisson}):
\begin{equation}\label{curlfreeL}
\nabla_{\bmath{x}} \times
\left(\ddot{\bmath{s}} + 2\frac{\dot a}{a}\dot{\bmath{s}} \right) =0 \,,
\end{equation}
\begin{equation}\label{poissonL}
\nabla_{\bmath{x}} \cdot
\left(\ddot{\bmath{s}} + 2\frac{\dot a}{a}\dot{\bmath{s}}
- \frac{1}{a^2} \frac{{\rm d} P}{{\rm d} \rho}
J^{-1} \nabla_{\bmath{x}} J \right)
= -4\pi G\rho_{\rm b} (J^{-1} -1) \,.
\end{equation}
The relation between $\nabla_{\bmath{x}}$ and
$\nabla_{\bmath{q}} \equiv \partial/\partial\bmath{q}$
is obtained from equation~(\ref{x=q+s}) as
\begin{equation}\label{partq-partx}
\frac{\partial}{\partial q_i} = \frac{\partial x_j}{\partial q_i}
                                \frac{\partial}{\partial x_j}
            = \left(\delta_{ji}+\frac{\partial s_j}{\partial q_i}\right)
                                \frac{\partial}{\partial x_j}
= \frac{\partial}{\partial x_i}+\frac{\partial s_j}{\partial q_i}
                                \frac{\partial}{\partial x_j} \,.
\end{equation}
Using this equation iteratively, we have
\begin{equation}
\frac{\partial}{\partial x_i} = \frac{\partial}{\partial q_i}
-\frac{\partial s_j}{\partial q_i}\frac{\partial}{\partial x_j}
  = \frac{\partial}{\partial q_i} - \frac{\partial s_j}{\partial q_i}
\left(\frac{\partial}{\partial q_j}-\frac{\partial s_k}{\partial q_j}
\frac{\partial}{\partial x_k}\right)
  = \frac{\partial}{\partial q_i}
-\frac{\partial s_j}{\partial q_i}\frac{\partial}{\partial q_j}
+\frac{\partial s_j}{\partial q_i}\frac{\partial s_k}{\partial q_j}
 \frac{\partial}{\partial x_k} = \cdots \,.
\label{partx-partq}
\end{equation}
The treatment is fully non-linear and exact so far.
Combining equations~(\ref{curlfreeL}), (\ref{poissonL}),
and (\ref{partx-partq}),
we can obtain perturbative solutions for $\bmath{s}$
up to any order in principle.
It should be emphasized that density $\rho$
is treated non-perturbatively because of equation~(\ref{exactrho}),
even if solutions for $\bmath{s}$ are obtained
in a perturbative manner.

\section{Perturbative approach in Lagrangian coordinates}

\subsection{Derivation of perturbation equations}

Let us proceed a perturbative approach in the Lagrangian description.
We write the displacement vector $\bmath{s}$ in a perturbative form
$\bmath{s}=\bmath{s}^{(1)}+\bmath{s}^{(2)}+\cdots$.
Superscripts $(1)$ and $(2)$ denote first-order and second-order
quantities in perturbative expansion with respect to amplitude $\epsilon$
of primordial fluctuations.
We make the perturbative expansion only for $\bmath{s}$,
and $\rho$ is not expanded.
Then we may expect relatively accurate description for $\rho$
by this formulation, even in non-linear regime.
In the perturbative expansion, equation~(\ref{curlfreeL}) gives
to first order,
\begin{equation}\label{curlfree1}
\nabla_{\bmath{q}} \times \left(\ddot{\bmath{s}}^{(1)}
+ 2\frac{\dot a}{a}\dot{\bmath{s}}^{(1)}\right) = 0 \,,
\end{equation}
and to second order,
\begin{equation}\label{curlfree2}
\left[\nabla_{\bmath{q}} \times \left(\ddot{\bmath{s}}^{(2)}
+ 2\frac{\dot a}{a}\dot{\bmath{s}}^{(2)}\right)\right]_i
= \epsilon_{ijk} s_{\ell\,,j}^{(1)} \left({\ddot s}_{k\,,\ell}^{(1)}
+ 2\frac{\dot a}{a}{\dot s}_{k\,,\ell}^{(1)} \right) \,,
\end{equation}
where $(\cdot)_{,i}$ denotes $\partial/\partial q_i$.

Next we consider equation~(\ref{poissonL}).
The Jacobian $J$ is expanded as
\[
J = 1 + \nabla_{\bmath{q}} \cdot \bmath{s}^{(1)}
      + \nabla_{\bmath{q}} \cdot \bmath{s}^{(2)}
      + \frac{1}{2}\left[
  (\nabla_{\bmath{q}} \cdot \bmath{s}^{(1)})^2
      - s_{i\,,j}^{(1)}s_{j\,,i}^{(1)} \right] + O(\epsilon^3) \,,
\]
and then the square of the `sound speed,' ${\rm d} P/{\rm d} \rho$,
can be written as
\[
\frac{{\rm d} P}{{\rm d} \rho}(\rho)
= \frac{{\rm d} P}{{\rm d} \rho}(\rho_{\rm b})
- \frac{{\rm d}^2 P}{{\rm d} \rho^2}(\rho_{\rm b}) \rho_{\rm b}
  \nabla_{\bmath{q}} \cdot \bmath{s}^{(1)} + O(\epsilon^2) \,.
\]
Thus we obtain to first order,
\begin{equation}\label{poisson1}
\nabla_{\bmath{q}} \cdot \left(
\ddot{\bmath{s}}^{(1)} + 2\frac{\dot a}{a}\dot{\bmath{s}}^{(1)}
-\frac{1}{a^2} \frac{{\rm d} P}{{\rm d} \rho}(\rho_{\rm b})
 \nabla_{\bmath{q}} (\nabla_{\bmath{q}} \cdot \bmath{s}^{(1)})
 \right) = 4\pi G\rho_{\rm b} \nabla_{\bmath{q}} \cdot \bmath{s}^{(1)} \,,
\end{equation}
and to second order,
\[
{\ddot s}_{i\,,i}^{(2)} + 2\frac{\dot a}{a}{\dot s}_{i\,,i}^{(2)}
- \frac{1}{a^2} \frac{{\rm d} P}{{\rm d} \rho}(\rho_{\rm b})
  \nabla_{\bmath{q}}^2 s_{i\,,i}^{(2)}
- s_{j\,,i}^{(1)} \left({\ddot s}_{i\,,j}^{(1)}
+ 2\frac{\dot a}{a}{\dot s}_{i\,,j}^{(1)} \right)
+\frac{1}{a^2} \frac{{\rm d} P}{{\rm d} \rho}(\rho_{\rm b})
\left( s_{i\,,ij}^{(1)}\nabla_{\bmath{q}}^2 s_{j}^{(1)}
+s_{i\,,jk}^{(1)}s_{j\,,ik}^{(1)}
+s_{i\,,j}^{(1)}\nabla_{\bmath{q}}^2 s_{j\,,i}^{(1)}
+2s_{i\,,j}^{(1)}s_{k\,,kij}^{(1)}
\right)
\]
\begin{equation}\label{poisson2}
+\frac{1}{a^2} \frac{{\rm d}^2 P}{{\rm d} \rho^2}(\rho_{\rm b}) \rho_{\rm b}
\left( s_{i\,,i}^{(1)}\nabla_{\bmath{q}}^2 s_{j\,,j}^{(1)}
+s_{i\,,ik}^{(1)}s_{j\,,jk}^{(1)} \right)
= 4\pi G\rho_{\rm b} \left[ s_{i\,,i}^{(2)} - \frac{1}{2}(s_{i\,,i}^{(1)})^2
- \frac{1}{2}s_{i\,,j}^{(1)}s_{j\,,i}^{(1)} \right] \,.
\end{equation}
In order to solve the perturbation equations,
it is convenient to decompose $\bmath{s}^{(1)}$
and $\bmath{s}^{(2)}$
into the longitudinal and the transverse parts in the form
\[
\bmath{s}^{(1)} = \nabla_{\bmath{q}} S + \bmath{S}^{\rm T} \,, \quad
\bmath{s}^{(2)} = \nabla_{\bmath{q}} \zeta + \bzeta^{\rm T} \,,
\]
where $S$ and $\zeta$ are respectively first-order and second-order
scalar functions,
and $\bmath{S}^{\rm T}$ and $\bzeta^{\rm T}$ satisfy
$\nabla_{\bmath{q}} \cdot \bmath{S}^{\rm T} =0$,
$\nabla_{\bmath{q}} \cdot \bzeta^{\rm T} =0$.
To note their physical meanings,
the first-order longitudinal and transverse parts are related
to linear density and vortical perturbations, respectively.
At the second-order level, however, such a simple interpretation
of the perturbation modes does not hold any more.
The first-order perturbation equations~(\ref{curlfree1})
and (\ref{poisson1}) then become
\begin{equation}
\nabla_{\bmath{q}} \times \left(\ddot{\bmath{S}^{\rm T}}
+ 2\frac{\dot a}{a}\dot{\bmath{S}^{\rm T}}
            \right) =0 \,,
\end{equation}
\begin{equation}
\nabla_{\bmath{q}}^2 \left(\ddot{S} + 2\frac{\dot a}{a}\dot{S}
- 4\pi G\rho_{\rm b} S
- \frac{1}{a^2} \frac{{\rm d} P}{{\rm d} \rho}(\rho_{\rm b})
\nabla_{\bmath{q}}^2 S \right) =0 \,.
\end{equation}
Under some adequate boundary conditions, these can be reduced as
\begin{equation}\label{trans1}
\ddot{\bmath{S}^{\rm T}} + 2\frac{\dot a}{a}\dot{\bmath{S}^{\rm T}} =0 \,,
\end{equation}
\begin{equation}\label{longit1}
\ddot{S} + 2\frac{\dot a}{a}\dot{S} - 4\pi G\rho_{\rm b} S
- \frac{1}{a^2} \frac{{\rm d} P}{{\rm d} \rho}(\rho_{\rm b})
\nabla_{\bmath{q}}^2 S =0 \,,
\end{equation}
which are obtained by Adler \& Buchert \shortcite{adler}.

The second-order perturbation equations~(\ref{curlfree2})
and (\ref{poisson2}) are also rewritten
in terms of the longitudinal and the transverse parts.
Equation~(\ref{curlfree2}) reads
\begin{equation}
\left[\nabla_{\bmath{q}} \times \left(\ddot{\bzeta^{\rm T}}
+ 2\frac{\dot a}{a} \dot{\bzeta^{\rm T}} \right)\right]_i
= 4\pi G\rho_{\rm b} \epsilon_{ijk} S^{\rm T}_{\ell\,,j} S_{,k\ell}
+ \frac{1}{a^2} \frac{{\rm d} P}{{\rm d} \rho}(\rho_{\rm b})
\epsilon_{ijk}(S_{,\ell j} + S^{\rm T}_{\ell\,,j})
\nabla_{\bmath{q}}^2 S_{,k\ell} \,.
\end{equation}
The curl of this equation gives
\begin{equation}\label{trans2}
-\nabla_{\bmath{q}}^2 \left( \ddot{\bzeta^{\rm T}}
+ 2\frac{\dot a}{a} \dot{\bzeta^{\rm T}} \right)
= \bmath{Q}^{\rm T} (\bmath{q},t) \,,
\end{equation}
where $\bmath{Q}^{\rm T}$ is a source term,
which is quadratic with respect to the first-order perturbations,
of the form
\begin{eqnarray*}
Q^{\rm T}_i (\bmath{q},t) &\equiv& 4\pi G\rho_{\rm b}
      \left(S^{\rm T}_{j\,,ik}S_{,jk}
          + S^{\rm T}_{j\,,i}\nabla_{\bmath{q}}^2 S_{,j}
          - \nabla_{\bmath{q}}^2 S^{\rm T}_j S_{,ij}
          - S^{\rm T}_{j\,,k}S_{,ijk} \right) \\
&& + \frac{1}{a^2} \frac{{\rm d} P}{{\rm d} \rho}(\rho_{\rm b})
     \left(S_{,ijk}\nabla_{\bmath{q}}^2 S_{,jk}
   + S_{,ij} \nabla_{\bmath{q}}^2 \nabla_{\bmath{q}}^2 S_{,j}
   - \nabla_{\bmath{q}}^2 S_{,j}\nabla_{\bmath{q}}^2 S_{,ij}
   - S_{,jk}\nabla_{\bmath{q}}^2 S_{,ijk} \right) \\
&& + \frac{1}{a^2} \frac{{\rm d} P}{{\rm d} \rho}(\rho_{\rm b})
   \left(S^{\rm T}_{j\,,ik}\nabla_{\bmath{q}}^2 S_{,jk}
   + S^{\rm T}_{j\,,i} \nabla_{\bmath{q}}^2 \nabla_{\bmath{q}}^2 S_{,j}
   - \nabla_{\bmath{q}}^2 S^{\rm T}_j \nabla_{\bmath{q}}^2 S_{,ij}
   - S^{\rm T}_{j\,,k} \nabla_{\bmath{q}}^2 S_{,ijk}
   \right) \,.
\end{eqnarray*}
Equation~(\ref{poisson2}) becomes
\begin{equation}\label{longit2}
\nabla_{\bmath{q}}^2 \left(
\ddot{\zeta} + 2\frac{\dot a}{a}\dot{\zeta} - 4\pi G\rho_{\rm b} \zeta
- \frac{1}{a^2} \frac{{\rm d} P}{{\rm d} \rho}(\rho_{\rm b})
\nabla_{\bmath{q}}^2 \zeta \right)
= Q(\bmath{q}, t) \,,
\end{equation}
where
\begin{eqnarray*}
Q(\bmath{q}, t) &\equiv& 2\pi G\rho_{\rm b}
   \left[S_{,ij}S_{,ij} - (\nabla_{\bmath{q}}^2 S)^2\right] \\
&& - \frac{1}{a^2} \frac{{\rm d} P}{{\rm d} \rho}(\rho_{\rm b})
   \left( \nabla_{\bmath{q}}^2 S_{,i}
          \nabla_{\bmath{q}}^2 S_{,i} + S_{,ijk} S_{,ijk}
   + 2S_{,ij} \nabla_{\bmath{q}}^2 S_{,ij} \right)
   - \frac{1}{a^2} \frac{{\rm d}^2 P}{{\rm d} \rho^2}(\rho_{\rm b})
   \rho_{\rm b} \left( \nabla_{\bmath{q}}^2 S
                       \nabla_{\bmath{q}}^2 \nabla_{\bmath{q}}^2 S
   + \nabla_{\bmath{q}}^2 S_{,i} \nabla_{\bmath{q}}^2 S_{,i} \right) \\
&& - \frac{1}{a^2} \frac{{\rm d} P}{{\rm d} \rho}(\rho_{\rm b})
   \left( S_{,ij}\nabla_{\bmath{q}}^2 S^{\rm T}_{i\,,j}
          + 2S_{,ijk} S^{\rm T}_{i\,,jk}
   + \nabla_{\bmath{q}}^2 S_{,i} \nabla_{\bmath{q}}^2 S^{\rm T}_i
   + 2\nabla_{\bmath{q}}^2 S_{,ij} S^{\rm T}_{i\,,j}
   \right) \\
&& - 2\pi G\rho_{\rm b} S^{\rm T}_{i\,,j} S^{\rm T}_{j\,,i}
   - \frac{1}{a^2} \frac{{\rm d} P}{{\rm d} \rho}(\rho_{\rm b})
    \left( S^{\rm T}_{i\,,j} \nabla_{\bmath{q}}^2 S^{\rm T}_{j\,,i}
    + S^{\rm T}_{i\,,jk} S^{\rm T}_{j\,,ik} \right) \,.
\end{eqnarray*}
We can easily confirm that equations~(\ref{trans2}) and (\ref{longit2})
are consistent with the second-order perturbation equations obtained
by Sasaki \& Kasai \shortcite{sasakasa} for the pressureless case.

\subsection{Solutions of perturbation equations}

Here we solve the perturbation equations in the presence of pressure effect.
We assume that the background universe is spatially flat one with
$a(t) = t^{2/3}$ and $\rho_{\rm b} = 1/(6\pi Gt^2) \propto a^{-3}$
for simplicity.
The first-order perturbation equation~(\ref{trans1})
for the transverse part has the same form as
in the pressureless case.
Thus we immediately find the solutions
\begin{equation}
\bmath{S}^{\rm T} \propto \ {\rm const.}\,, \ t^{-1/3} \,.
\end{equation}
For the longitudinal part, 
the Fourier transform of equation~(\ref{longit1})
with respect to the Lagrangian coordinates yields
\begin{equation}\label{ddothatS}
\ddot{\widehat{S}} + 2\frac{\dot a}{a}\dot{\widehat{S}}
- 4\pi G\rho_{\rm b} \widehat{S}
+ \frac{1}{a^2} \frac{{\rm d} P}{{\rm d} \rho}(\rho_{\rm b})
|\bmath{K}|^2 \widehat{S} =0 \,,
\end{equation}
where $\widehat{(\cdot)}$ denotes a Fourier component,
and $\bmath{K}$ is a wavenumber vector
associated with the Lagrangian coordinates $\bmath{q}$.
It should be noted that the form of equation~(\ref{ddothatS})
is similar to that of an equation for the density contrast
$\delta_{\rm lin}$ in the Eulerian linear theory.
Actually it reads
\begin{equation}\label{ddotdeltalin}
\frac{\partial^2 \delta_{\rm lin}(\bmath{k},t)}{\partial t^2}
+ 2\frac{\dot a}{a}
  \frac{\partial \delta_{\rm lin}(\bmath{k},t)}{\partial t}
- 4\pi G\rho_{\rm b} \delta_{\rm lin}(\bmath{k},t)
+ \frac{1}{a^2} \frac{{\rm d} P}{{\rm d} \rho} (\rho_{\rm b})
  |\bmath{k}|^2 \delta_{\rm lin}(\bmath{k},t) =0 \,,
\end{equation}
where $\bmath{k}$ denotes a wavenumber vector
associated with the Eulerian coordinates $\bmath{x}$.
As an example, assuming a polytropic equation of state
$P=\kappa\rho^{\gamma}$
with a constant $\kappa$ and a polytropic index $\gamma$,
the solutions of equation~(\ref{ddotdeltalin}) are \cite{weinberg}
\begin{equation}\label{deltalin-k}
\delta_{\rm lin}(\bmath{k},t) \propto \left\{
\begin{array}{ll}
t^{-1/6} J_{\pm \nu} \left( A |\bmath{k}| t^{-\gamma+4/3} \right) \qquad
 & \mbox{for} \quad \gamma \ne \frac{4}{3} \,, \\
t^{-1/6 \pm \sqrt{25/36 - B|\bmath{k}|^2}}
 & \mbox{for} \quad \gamma = \frac{4}{3} \,,
\end{array}
\right.
\end{equation}
where $\nu = 5/(8-6\gamma)$,
$J_{\pm \nu}$ is the Bessel function of order $\pm \nu$, and
\[
A \equiv \frac{3\sqrt{\kappa\gamma}(6\pi G)^{(1-\gamma)/2}}{|4-3\gamma|}\,,
\quad
B \equiv \frac{4}{3}\kappa (6\pi G)^{-1/3} \,.
\]
Note that the above solutions include wavenumbers of fluctuations,
whereas the solutions for the pressureless matter do not.
We then find solutions of equation~(\ref{ddothatS})
with the help of the known results
for the density contrast in the Eulerian linear theory.
Hence, in the case of the polytropic equation of state $P=\kappa\rho^{\gamma}$
and if $\nu=5/(8-6\gamma)$ is not an integer, we obtain
general solutions for $\widehat{S}(\bmath{K},t)$ as
\begin{equation}\label{hatS}
\widehat{S}(\bmath{K},t) =
D^+(\bmath{K},t) C^+(\bmath{K}) + D^-(\bmath{K},t) C^-(\bmath{K}) \,,
\end{equation}
where $D^{\pm}(\bmath{K},t)$ are provided,
by replacing $\bmath{k}$ with $\bmath{K}$
in equation~(\ref{deltalin-k}), in the form
\begin{equation}
D^{\pm}(\bmath{K},t) = \left\{
\begin{array}{ll}
t^{-1/6} J_{\pm \nu} \left( A |\bmath{K}| t^{-\gamma+4/3} \right) \qquad
 & \mbox{for} \quad \gamma \ne \frac{4}{3} \,, \\
t^{-1/6 \pm \sqrt{25/36 - B|\bmath{K}|^2}}
 & \mbox{for} \quad \gamma = \frac{4}{3} \,,
\end{array}
\right.
\end{equation}
and $C^{\pm}(\bmath{K})$ are determined by initial conditions.
Notice that $\bmath{K}$ is the Lagrangian wavenumber,
which is different from the Eulerian wavenumber $\bmath{k}$.

Next we consider solutions of the second-order perturbation
equations~(\ref{trans2}) and (\ref{longit2}).
The Fourier transform of equations~(\ref{trans2}) and
(\ref{longit2}) gives
\begin{equation}
|\bmath{K}|^2 \left( \ddot{\widehat{\bzeta^{\rm T}}}
+ 2\frac{\dot a}{a}\dot{\widehat{\bzeta^{\rm T}}} \right)
= \widehat{\bmath{Q}^{\rm T}} (\bmath{K},t) \,,
\end{equation}
\begin{equation}
-|\bmath{K}|^2 \left( \ddot{\widehat{\zeta}}
+ 2\frac{\dot a}{a}\dot{\widehat{\zeta}}
- 4\pi G\rho_{\rm b} \widehat{\zeta}
+ \frac{1}{a^2} \frac{{\rm d} P}{{\rm d} \rho}(\rho_{\rm b})
|\bmath{K}|^2 \widehat{\zeta} \right)
= \widehat{Q}(\bmath{K},t) \,.
\end{equation}
The solutions are formally written as
\begin{equation}
\widehat{\bzeta^{\rm T}}(\bmath{K},t) = \frac{1}{|\bmath{K}|^2}
\int^t {\rm d} t' \, G^{\rm T}(t,t')
\,\widehat{\bmath{Q}^{\rm T}}(\bmath{K},t') \,,
\end{equation}
\begin{equation}
\widehat{\zeta}(\bmath{K},t) = -\frac{1}{|\bmath{K}|^2}
\int^t {\rm d} t' \, G(\bmath{K},t,t')
\,\widehat{Q}(\bmath{K},t') \,,
\end{equation}
by using the Green functions $G^{\rm T}(t,t')$ and $G(\bmath{K},t,t')$.
The Green function $G^{\rm T}(t,t')$ does not depend on
an equation of state and is given as
\begin{equation}
G^{\rm T}(t,t') = 3(t'-t^{-1/3}t'^{4/3}) \,,
\end{equation}
while the $G(\bmath{K},t,t')$ depends on an equation of state.
Under the assumption $P=\kappa\rho^{\gamma}$,
if $\gamma \ne 4/3$ and $\nu=5/(8-6\gamma)$ is not an integer,
we have
\begin{eqnarray}
G(\bmath{K},t,t') &=& -\frac{\pi}{2\sin\nu\pi}
                       \left(-\gamma+\frac{4}{3}\right)^{-1}
 t^{-1/6}t'^{7/6} \biggl[ J_{-\nu}(A|\bmath{K}|t^{-\gamma+4/3})
                          J_{\nu}(A|\bmath{K}|t'^{-\gamma+4/3})\nonumber \\
 &&
 \qquad
 -J_{\nu}(A|\bmath{K}|t^{-\gamma+4/3})
  J_{-\nu}(A|\bmath{K}|t'^{-\gamma+4/3}) \biggr] \,,
\end{eqnarray}
and if $\gamma=4/3$,
\begin{equation}
G(\bmath{K},t,t') = -\frac{1}{2}
                     \left(\frac{25}{36}-B|\bmath{K}|^2\right)^{-1/2}
 t^{-1/6}t'^{7/6} \left(
 t^{-\sqrt{25/36-B|\bmath{K}|^2}}
 t'^{\sqrt{25/36-B|\bmath{K}|^2}}
 -t^{\sqrt{25/36-B|\bmath{K}|^2}} t'^{-\sqrt{25/36-B|\bmath{K}|^2}}
                  \right) \,.
\end{equation}
In order to present explicit form of the second-order solutions,
we must compute the Fourier-transformed source terms
$\widehat{\bmath{Q}^{\rm T}}$ and $\widehat{Q}$.
Hereafter we neglect the first-order transverse part $\bmath{S}^{\rm T}$
in $\widehat{\bmath{Q}^{\rm T}}$ and $\widehat{Q}$ for simplicity.
This is equivalent to give no attention to effect of vorticity
in the second-order solutions.
They are written in the following convolution form:
\begin{eqnarray}
\widehat{\bmath{Q}^{\rm T}}(\bmath{K},t) &=&
 -\frac{i}{(2\pi)^3} \frac{1}{a^2}
\frac{{\rm d} P}{{\rm d} \rho}(\rho_{\rm b})
\int^{\infty}_{-\infty} {\rm d}^3 \bmath{K}'
\,\widehat{S}(\bmath{K}',t)
\,\widehat{S}(\bmath{K}-\bmath{K}',t)
\,|\bmath{K}-\bmath{K}'|^2 \,\bmath{K}' \cdot(\bmath{K}-\bmath{K}')
\nonumber \\
 &&
\ \cdot \ 
\Bigl[ \bmath{K}' \left(\bmath{K}' \cdot(\bmath{K}-\bmath{K}')\right)
 + \bmath{K}'|\bmath{K}-\bmath{K}'|^2
 - (\bmath{K}-\bmath{K}') |\bmath{K}'|^2 - (\bmath{K}-\bmath{K}')
\left(\bmath{K}' \cdot(\bmath{K}-\bmath{K}')\right)
\Bigr] \,,
\end{eqnarray}
\begin{eqnarray}
\widehat{Q}(\bmath{K},t) &=& \frac{1}{(2\pi)^3}
\int^{\infty}_{-\infty} {\rm d}^3 \bmath{K}'
\,\widehat{S}(\bmath{K}',t)
\,\widehat{S}(\bmath{K}-\bmath{K}',t)
\,\biggl[ 2\pi G\rho_{\rm b} \left[
\left(\bmath{K}' \cdot(\bmath{K}-\bmath{K}')\right)^2
 - |\bmath{K}'|^2 |\bmath{K}-\bmath{K}'|^2 \right] \nonumber \\
 &&
+ \frac{1}{a^2} \frac{{\rm d} P}{{\rm d} \rho}(\rho_{\rm b})
\Bigl[ |\bmath{K}'|^2 |\bmath{K}-\bmath{K}'|^2
       \,\bmath{K}' \cdot(\bmath{K}-\bmath{K}')
+ \left(\bmath{K}' \cdot(\bmath{K}-\bmath{K}')\right)^3
+ 2|\bmath{K}-\bmath{K}'|^2 \left(\bmath{K}' \cdot(\bmath{K}-\bmath{K}')\right)^2
\Bigr] \nonumber \\
 && +
\frac{1}{a^2} \frac{{\rm d}^2 P}{{\rm d} \rho^2}(\rho_{\rm b})\rho_{\rm b}
\left[ |\bmath{K}'|^2 |\bmath{K}-\bmath{K}'|^4
     + |\bmath{K}'|^2 |\bmath{K}-\bmath{K}'|^2
\,\bmath{K}' \cdot(\bmath{K}-\bmath{K}')
\right]
\biggr] .
\end{eqnarray}
Using the first-order solution~(\ref{hatS}),
we obtain
\begin{eqnarray}
\widehat{\bzeta^{\rm T}}(\bmath{K},t) &=& -\frac{i}{(2\pi)^3}
\frac{1}{|\bmath{K}|^2}
\int^{\infty}_{-\infty} {\rm d}^3 \bmath{K}'
\,E^{\rm T}(\bmath{K},\bmath{K}',t)
\Bigl(C^+(\bmath{K}') C^+(\bmath{K}-\bmath{K}')
+ C^+(\bmath{K}') C^-(\bmath{K}-\bmath{K}') \nonumber \\
 &&
\quad
+ C^-(\bmath{K}') C^+(\bmath{K}-\bmath{K}')
+ C^-(\bmath{K}') C^-(\bmath{K}-\bmath{K}')\Bigr)
\ |\bmath{K}-\bmath{K}'|^2
\ \bmath{K}' \cdot(\bmath{K}-\bmath{K}') \nonumber \\
 &&
\Bigl[ \bmath{K}' \left(\bmath{K}' \cdot(\bmath{K}-\bmath{K}')\right) 
 + \bmath{K}' |\bmath{K}-\bmath{K}'|^2
 - (\bmath{K}-\bmath{K}') |\bmath{K}'|^2
 - (\bmath{K}-\bmath{K}')
\left(\bmath{K}' \cdot(\bmath{K}-\bmath{K}')\right)
\Bigr] \,,
\label{hatzetaT}
\end{eqnarray}
\begin{eqnarray}
\widehat{\zeta}(\bmath{K},t) &=& -\frac{1}{(2\pi)^3} \frac{1}{|\bmath{K}|^2}
\int^{\infty}_{-\infty} {\rm d}^3 \bmath{K}'
\Bigl(C^+(\bmath{K}') C^+(\bmath{K}-\bmath{K}')
    + C^+(\bmath{K}') C^-(\bmath{K}-\bmath{K}') \nonumber \\
 &&
\qquad
+ C^-(\bmath{K}') C^+(\bmath{K}-\bmath{K}')
+ C^-(\bmath{K}') C^-(\bmath{K}-\bmath{K}') \Bigr)
\ \biggl[ E(\bmath{K},\bmath{K}',t)
\left[ \left(\bmath{K}' \cdot(\bmath{K}-\bmath{K}')\right)^2
 - |\bmath{K}'|^2 |\bmath{K}-\bmath{K}'|^2 \right] \nonumber \\
 &&
\qquad
+ F_1(\bmath{K},\bmath{K}',t)
\Bigl[ |\bmath{K}'|^2 |\bmath{K}-\bmath{K}'|^2
\,\bmath{K}' \cdot(\bmath{K}-\bmath{K}')
+ \left(\bmath{K}' \cdot(\bmath{K}-\bmath{K}')\right)^3
+ 2|\bmath{K}-\bmath{K}'|^2
\left(\bmath{K}' \cdot(\bmath{K}-\bmath{K}')\right)^2
\Bigr] \nonumber \\
 &&
\qquad
+ F_2(\bmath{K},\bmath{K}',t)
\left[ |\bmath{K}'|^2 |\bmath{K}-\bmath{K}'|^4
     + |\bmath{K}'|^2 |\bmath{K}-\bmath{K}'|^2
\,\bmath{K}' \cdot(\bmath{K}-\bmath{K}')
\right]
\biggr] \,,
\label{hatzeta}
\end{eqnarray}
where time-dependent factors are given as
\begin{eqnarray}
E^{\rm T}(\bmath{K},\bmath{K}',t) &=& \int^t \frac{{\rm d} t'}{a^2(t')}
\frac{{\rm d} P}{{\rm d} \rho}(\rho_{\rm b}(t')) \,G^{\rm T}(t,t')
\,\Bigl(D^+(\bmath{K}',t') D^+(\bmath{K}-\bmath{K}',t')
+ D^+(\bmath{K}',t') D^-(\bmath{K}-\bmath{K}',t')
\nonumber \\
 &&
\quad
+ D^-(\bmath{K}',t') D^+(\bmath{K}-\bmath{K}',t')
+ D^-(\bmath{K}',t') D^-(\bmath{K}-\bmath{K}',t')\Bigr) \,,
\label{ET(K,K',t)}
\end{eqnarray}
\begin{eqnarray}
E(\bmath{K},\bmath{K}',t) &=& \int^t {\rm d} t' \, 2\pi G\rho_{\rm b}(t')
\,G(\bmath{K},t,t')
\,\Bigl(D^+(\bmath{K}',t') D^+(\bmath{K}-\bmath{K}',t')
+ D^+(\bmath{K}',t') D^-(\bmath{K}-\bmath{K}',t')
\nonumber \\
 &&
\quad
+ D^-(\bmath{K}',t') D^+(\bmath{K}-\bmath{K}',t')
+ D^-(\bmath{K}',t') D^-(\bmath{K}-\bmath{K}',t')\Bigr) \,,
\end{eqnarray}
\begin{eqnarray}
F_1(\bmath{K},\bmath{K}',t) &=& \int^t \frac{{\rm d} t'}{a^2(t')}
\frac{{\rm d} P}{{\rm d} \rho}(\rho_{\rm b}(t')) \,G(\bmath{K},t,t')
\,\Bigl(D^+(\bmath{K}',t') D^+(\bmath{K}-\bmath{K}',t')
+ D^+(\bmath{K}',t') D^-(\bmath{K}-\bmath{K}',t')
\nonumber \\
 &&
\quad
+ D^-(\bmath{K}',t') D^+(\bmath{K}-\bmath{K}',t')
+ D^-(\bmath{K}',t') D^-(\bmath{K}-\bmath{K}',t')\Bigr) \,,
\end{eqnarray}
\begin{eqnarray}
F_2(\bmath{K},\bmath{K}',t) &=& \int^t \frac{{\rm d} t'}{a^2(t')}
\frac{{\rm d}^2 P}{{\rm d} \rho^2}(\rho_{\rm b}(t')) \rho_{\rm b}(t')
\,G(\bmath{K},t,t')
\,\Bigl(D^+(\bmath{K}',t') D^+(\bmath{K}-\bmath{K}',t')
+ D^+(\bmath{K}',t') D^-(\bmath{K}-\bmath{K}',t')
\nonumber \\
 &&
\quad
+ D^-(\bmath{K}',t') D^+(\bmath{K}-\bmath{K}',t')
+ D^-(\bmath{K}',t') D^-(\bmath{K}-\bmath{K}',t')\Bigr) \,.
\label{F2(K,K',t)}
\end{eqnarray}
The convolution in the solutions~(\ref{hatzetaT}) and (\ref{hatzeta})
represents mode couplings in $\bmath{K}$-space,
which inevitably occur at second order due to non-linearity.
Although it is cumbersome to perform the integration
in equations~(\ref{ET(K,K',t)})--(\ref{F2(K,K',t)}) in general,
it is easy to do it if an equation of state is $P=\kappa\rho^{4/3}$.
In this case, if only the $D^+(\bmath{K}',t) D^+(\bmath{K}-\bmath{K}',t)$
part is considered,
equations~(\ref{ET(K,K',t)})--(\ref{F2(K,K',t)}) become
\begin{equation}
E^{\rm T}(\bmath{K},\bmath{K}',t)
= \frac{B t^{-1/3 + b(\bmath{K}') + b(\bmath{K}-\bmath{K}')}}
       {\left[-\frac{1}{3} + b(\bmath{K}') + b(\bmath{K}-\bmath{K}')\right] 
        \left[b(\bmath{K}') + b(\bmath{K}-\bmath{K}')\right]} \,,
\end{equation}
\begin{equation}
E(\bmath{K},\bmath{K}',t)
= \frac{t^{-1/3 + b(\bmath{K}') + b(\bmath{K}-\bmath{K}')}}
{3 \left[-\frac{1}{6} - b(\bmath{K}) + b(\bmath{K}')
         + b(\bmath{K}-\bmath{K}')\right]
   \left[-\frac{1}{6} + b(\bmath{K}) + b(\bmath{K}')
         + b(\bmath{K}-\bmath{K}')\right]} \,,
\end{equation}
\begin{equation}
F_1(\bmath{K},\bmath{K}',t)
= \frac{B t^{-1/3 + b(\bmath{K}') + b(\bmath{K}-\bmath{K}')}}
       {\left[-\frac{1}{6} - b(\bmath{K}) + b(\bmath{K}')
              + b(\bmath{K}-\bmath{K}')\right]
        \left[-\frac{1}{6} + b(\bmath{K}) + b(\bmath{K}')
              + b(\bmath{K}-\bmath{K}')\right]} \,,
\end{equation}
\begin{equation}
F_2(\bmath{K},\bmath{K}',t)
= \frac{B t^{-1/3 + b(\bmath{K}') + b(\bmath{K}-\bmath{K}')}}
{3 \left[-\frac{1}{6} - b(\bmath{K}) + b(\bmath{K}')
         + b(\bmath{K}-\bmath{K}')\right]
   \left[-\frac{1}{6} + b(\bmath{K}) + b(\bmath{K}')
         + b(\bmath{K}-\bmath{K}')\right]} \,,
\end{equation}
where $b(\bmath{K}) = \sqrt{25/36 - B|\bmath{K}|^2}$.
Note that all these factors have the same temporal dependence,
$t^{-1/3 + b(\bmath{K}') + b(\bmath{K}-\bmath{K}')}$.
Of course, it is not a general property of the second-order solutions,
but $F_1(\bmath{K},\bmath{K}',t)$ and $F_2(\bmath{K},\bmath{K}',t)$
always have the same temporal dependence
as long as an equation of state is of the form $P=\kappa\rho^{\gamma}$.

\section{Illustration in a one-dimensional model}

In this section, we present illustrative examples
of computation by the Lagrangian perturbation theory
formulated in the previous section.
We compute power spectra of density perturbations
by the Eulerian linear theory, the Lagrangian first-order and second-order
approximations in a one-dimensional model and then clarify difference
between the Eulrian and the Lagrangian approximations.
In linear regime $|\delta| \ll 1$, the Eulerian and the Lagrangian
approximations give the same results.
However, when these approximations are extrapolated into non-linear regime,
the results given by them do not always coincide.

In the pressureless case, the first-order approximation
(i.e. the Zel'dovich approximation)
coincides with an exact solution in a one-dimensional model.
Although this does not hold in the presence of pressure in general,
we discuss that the Lagrangian perturbative approximations
may provide nearly exact description
in weakly non-linear regime $|\delta| \la 1$,
by consulting difference between the first-order and the second-order
approximations.
Of course, we cannot say that one-dimensional examples are realistic,
but they are instructive to show advantages and features of non-linearity
which the Lagrangian perturbation theory involves.

\subsection{Equations and perturbative solutions in a one-dimensional model}

First we present basic equations and perturbative solutions
in a one-dimensional model.
In the Eulerian linear approximation,
the density contrast satisfies equation~(\ref{ddotdeltalin}), that is
\begin{equation}\label{ddotdeltalin1D}
\frac{\partial^2 \delta_{\rm lin}(k,t)}{\partial t^2}
+ 2\frac{\dot a}{a}
  \frac{\partial \delta_{\rm lin}(k,t)}{\partial t}
- 4\pi G\rho_{\rm b} \delta_{\rm lin}(k,t)
+ \frac{1}{a^2} \frac{{\rm d} P}{{\rm d} \rho} (\rho_{\rm b})
  k^2 \delta_{\rm lin}(k,t) =0 \,,
\end{equation}
where $k$ is the first component of the Eulerian wavenumber vector $\bmath{k}$.
We find from this equation that density perturbations
whose wavenumbers are smaller than
\[
k_{\rm J} \equiv \left( \frac{4\pi G\rho_{\rm b}a^2}
                             {{\rm d}P/{\rm d}\rho}
                 \right)^{1/2}
\]
will grow to form inhomogeneous structures,
and those whose wavenumbers are larger than $k_{\rm J}$
will decay with acoustic oscillations (the Jeans condition).
In particular, we immediately see the behaviour of density perturbations
in the case $P=\kappa\rho^{4/3}$,
where the solutions of equation~(\ref{ddotdeltalin1D}) are
\begin{equation}
\delta_{\rm lin}(k,t) \propto t^{-1/6 \pm \sqrt{25/36 -Bk^2}} \,,
\end{equation}
and $k_{\rm J} = \sqrt{2/(3B)}$.
If $k<k_{\rm J}$, one of the solutions becomes a growing solution,
whereas if $k>k_{\rm J}$, both of the solutions are decaying ones.

The relation between the Eulerian and the Lagrangian coordinates,
equation~(\ref{x=q+s}), in a one-dimensional model can be written as
\begin{equation}
\left\{
\begin{array}{ll}
x_1 & = q_1 + s_1(q_1,t) \,, \\
x_2 & = q_2 \,, \\
x_3 & = q_3 \,,
\end{array}
\right.
\end{equation}
and energy density, equation~(\ref{exactrho}), is then
\begin{equation}\label{rho1dim}
\rho(q_1,t) = \frac{\rho_{\rm b}(t)}{1 + s_{1,1}(q_1,t)} \,,
\end{equation}
because the Jacobian $J = 1+s_{1,1}$.
The relation between $\nabla_{\bmath{x}}$ and $\nabla_{\bmath{q}}$,
equation~(\ref{partq-partx}), becomes
\begin{equation}
\frac{\partial}{\partial x_1} = \frac{1}{J}
                                \frac{\partial}{\partial q_1} \,, \quad
\frac{\partial}{\partial x_2} = \frac{\partial}{\partial q_2} \,, \quad
\frac{\partial}{\partial x_3} = \frac{\partial}{\partial q_3} \,.
\end{equation}
Hence from equation~(\ref{poissonL}) we have
\begin{equation}
\frac{1}{J} \frac{\partial}{\partial q_1}
\left( \ddot{s_1} + 2\frac{\dot a}{a}\dot{s_1}
     - \frac{1}{a^2} \frac{{\rm d} P}{{\rm d} \rho} J^{-2} J_{,1}
\right) = -4\pi G\rho_{\rm b} (J^{-1} -1) \,,
\end{equation}
which can be reduced as
\begin{equation}\label{ddots1dPdrho}
\ddot{s_1} + 2\frac{\dot a}{a}\dot{s_1} - 4\pi G\rho_{\rm b} s_1
- \frac{1}{a^2} \frac{{\rm d} P}{{\rm d} \rho}
  \frac{s_{1,11}}{(1+s_{1,1})^2} = 0 \,,
\end{equation}
by imposing appropriate boundary conditions.
If we assume that an equation of state is of the form $P=\kappa\rho^{\gamma}$,
we obtain by using equation~(\ref{rho1dim}),
\begin{equation}\label{ddots1poly}
\ddot{s_1} + 2\frac{\dot a}{a}\dot{s_1} - 4\pi G\rho_{\rm b} s_1
- \frac{\kappa\gamma\rho_{\rm b}^{\gamma-1}}{a^2}
  \frac{s_{1,11}}{(1+s_{1,1})^{1+\gamma}} = 0 \,.
\end{equation}
It seems to be difficult to solve equation~(\ref{ddots1dPdrho}) or
equation~(\ref{ddots1poly}) exactly in general,
although G\"{o}tz \shortcite{goetz} solved it
in the case $P \propto \rho$ without cosmic expansion.
Then we consider their solutions in a perturbative manner
and adopt the perturbation solutions obtained in the previous section.
Note that the displacement vector $\bmath{s} = (s_1,0,0)$ consists
only of the longitudinal parts ($S_{,1}$, $\zeta_{,1}$, $\ldots$),
because the transverse parts ($\bmath{S}^{\rm T}$, $\bmath{\zeta}^{\rm T}$, $\ldots$)
vanish in a one-dimensional model.
The perturbation solutions~(\ref{hatS}) and (\ref{hatzeta}) become
\begin{equation}
\widehat{S}(K,t) = D^+(K,t) C^+(K) + D^-(K,t) C^-(K) \,,
\end{equation}
\begin{eqnarray}
\widehat{\zeta}(K,t) &=& -\frac{1}{2\pi K}
\int^{\infty}_{-\infty} {\rm d} K'
\left(C^+(K') C^+(K-K') + C^+(K') C^-(K-K')
+ C^-(K') C^+(K-K') + C^-(K') C^-(K-K')\right) \nonumber \\
 &&
\quad \cdot \ 
\left[ 2F_1(K,K',t) + F_2(K,K',t) \right]
K'^2 (K-K')^3 \,,
\end{eqnarray}
where $K$ is the first component of the Lagrangian wavenumber vector $\bmath{K}$.
The part proportional to $E(\bmath{K},\bmath{K}',t)$ in the second-order
solutions~(\ref{hatzeta}) does not appear in the above expression
because it vanishes in a one-dimensional model.

In order to simplify the perturbation solutions further,
let us consider the case where an equation of state is $P=\kappa\rho^{4/3}$.
Although the validity of this assumption is not clarified,
it would be useful to understand features of the perturbation theory
we have formulated.
The temporal factors are computed as
\begin{equation}
D^{\pm}(K,t) = t^{-1/6 \pm b(K)} \,,
\end{equation}
\begin{equation}
2 F_1 (K,K',t) + F_2 (K,K',t)
= \frac{7}{3} \frac{B t^{-1/3 + b(K') + b(K-K')}}
  {\left[-\frac{1}{6} - b(K) + b(K') + b(K-K')\right]
   \left[-\frac{1}{6} + b(K) + b(K') + b(K-K')\right]} \,,
\end{equation}
where we again take into account only the $D^+(K')D^+(K-K')$ part
to compute $2F_1+F_2$.

\subsection{Initial conditions}

We consider setting of initial conditions for illustration.
Initial conditions for two independent physical quantities are needed
to determine $C^{\pm}(K)$.
Here we impose initial conditions for the density contrast $\delta$
and the peculiar velocity $\bmath{v}$,
whose initial values are denoted by $\delta_{\rm in}$ and $\bmath{v}_{\rm in}$, respectively.
 For comparison with the pressureless case,
we take $\delta_{\rm in}$ and $\bmath{v}_{\rm in}$
as those given by the Zel'dovich approximation,
which is a subclass of the first-order approximation for pressureless fluid
and becomes an exact solution in a one-dimensional model.
By this setting, the $C^{\pm}(K)$ are expressed in terms of only $\delta_{\rm in}$,
because the Zel'dovich approximation includes just one arbitrary
spatial function, through which we can make a relation between $\delta_{\rm in}$ and $\bmath{v}_{\rm in}$.
The one-dimensional Zel'dovich approximation is written as
\begin{equation}
\left\{
\begin{array}{ll}
x_1 & = q_1 + t^{2/3} \Psi_{,1}(q_1) \,, \\
x_2 & = q_2 \,, \\
x_3 & = q_3 \,,
\end{array}
\right.
\end{equation}
\begin{equation}
\rho(q_1,t) = \frac{\rho_{\rm b}(t)}
                   {1+t^{2/3}\Psi_{,11}(q_1)} \,,
\end{equation}
where $\Psi(q_1)$ is an arbitrary spatial function, describing initial inhomogeneity.
 From these equations, we have
\begin{equation}\label{deltainitialZA}
\delta_{\rm in}(q_1)
= \frac{1}{1+t^{2/3}\Psi_{,11}(q_1)}-1 \biggl|_{t=t_{\rm in}}
\simeq -\Psi_{,11}(q_1) \,,
\end{equation}
\begin{equation}\label{pecinitialZA}
\bmath{v}_{\rm in}(q_1)
= \left((2/3) t^{1/3} \Psi_{,1}(q_1), 0,0 \right) \biggl|_{t=t_{\rm in}}
= \left((2/3) \Psi_{,1}(q_1), 0,0 \right) \,,
\end{equation}
where we define an initial time $t_{\rm in} \equiv 1$.
On the other hand, the first-order solution in the case $P=\kappa\rho^{4/3}$ gives
\begin{equation}\label{deltain43}
\widehat{\delta_{\rm in}}(K) = K^2 (C^+(K) + C^-(K)) \,,
\end{equation}
\begin{equation}\label{vin43}
\widehat{\bmath{v}_{\rm in}}(K) = \left(
  iK \left[ \left(-1/6+b(K)\right) C^+(K)
           +\left(-1/6-b(K)\right) C^-(K) \right],0,0 \right) \,.
\end{equation}
Comparing equations~(\ref{deltainitialZA}), (\ref{pecinitialZA}),
(\ref{deltain43}), and (\ref{vin43}), we obtain
\begin{equation}\label{Cpm43}
C^{\pm}(K) = \frac{\widehat{\delta_{\rm in}}(K)}{2 K^2}
             \left( 1 \pm \frac{5}{6b(K)} \right) \,.
\end{equation}
Thus $C^{\pm}(K)$ are completely determined
if $\widehat{\delta_{\rm in}}(K)$ is provided
in some appropriate manner.
In our illustration, we choose
$\widehat{\delta_{\rm in}}(K) = |\widehat{\delta_{\rm in}}(K)|\exp(i\phi_K)$
so that $|\widehat{\delta_{\rm in}}(K)|^2 \propto |K|^n$,
where $n$ is a spectral index,
and the phases $\phi_K$ are randomly distributed on the interval $[0,2\pi]$.
This choice is a simplification of the Gaussian statistics
for initial density perturbations $\delta_{\rm in}(q_1)$
in real space \cite{coles},
which is usually adopted in the study of large-scale structure formation.

\subsection{Evolution of power spectra of density perturbations}

\begin{figure}
\begin{minipage}{.50\linewidth}
\leavevmode
\epsfysize=85mm
\centerline{\epsfbox{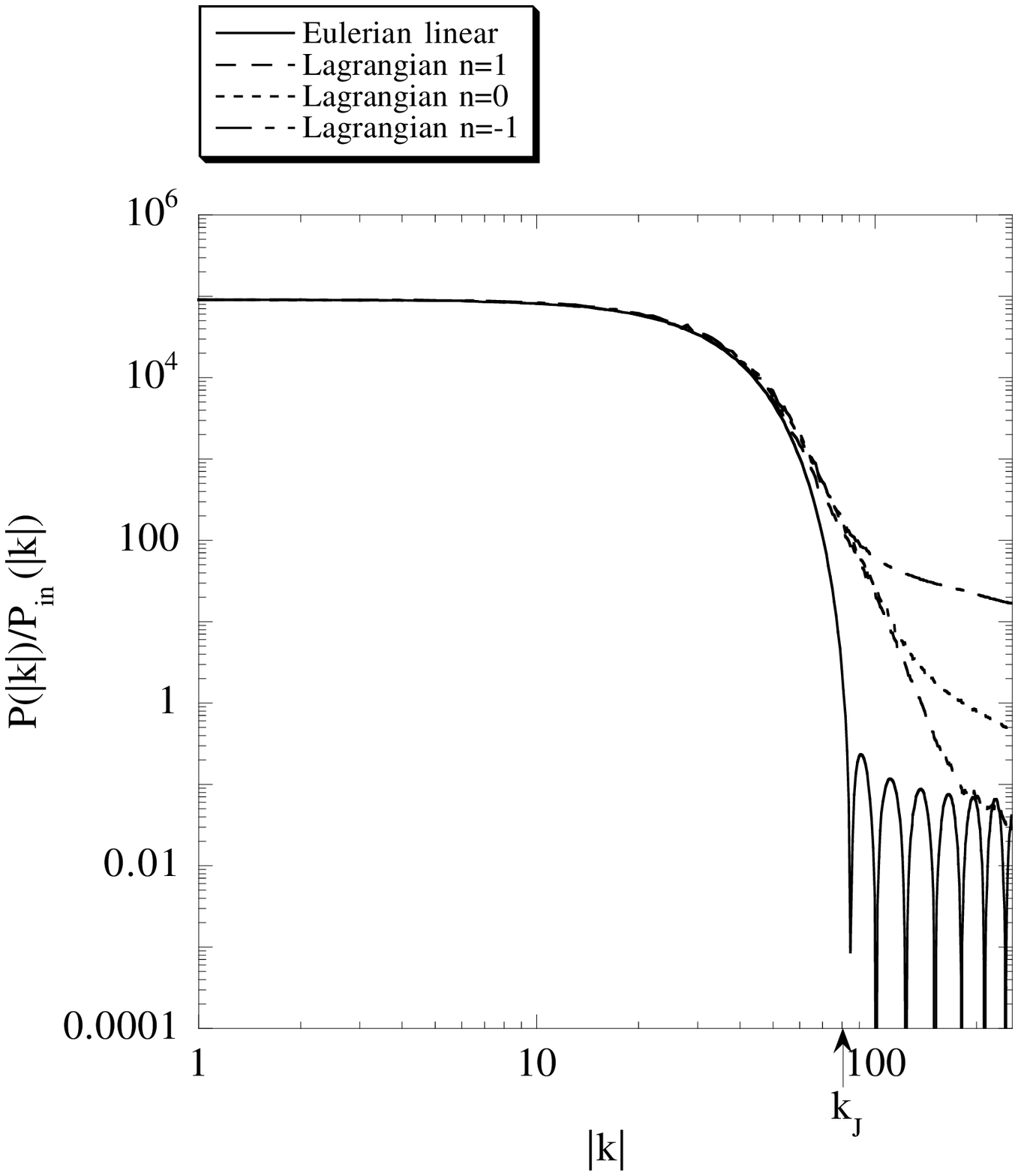}}
\caption{The `transfer function' of density perturbations at $a=300$
         computed from the Eulerian linear theory
         and the Lagrangian first-order approximations.
         It does not depend on the initial conditions
         in the Eulerian linear theory,
         but does in the Lagrangian approximation.}
\label{trans0300}
\end{minipage}
\begin{minipage}{.45\linewidth}
\leavevmode
\epsfysize=85mm
\centerline{\epsfbox{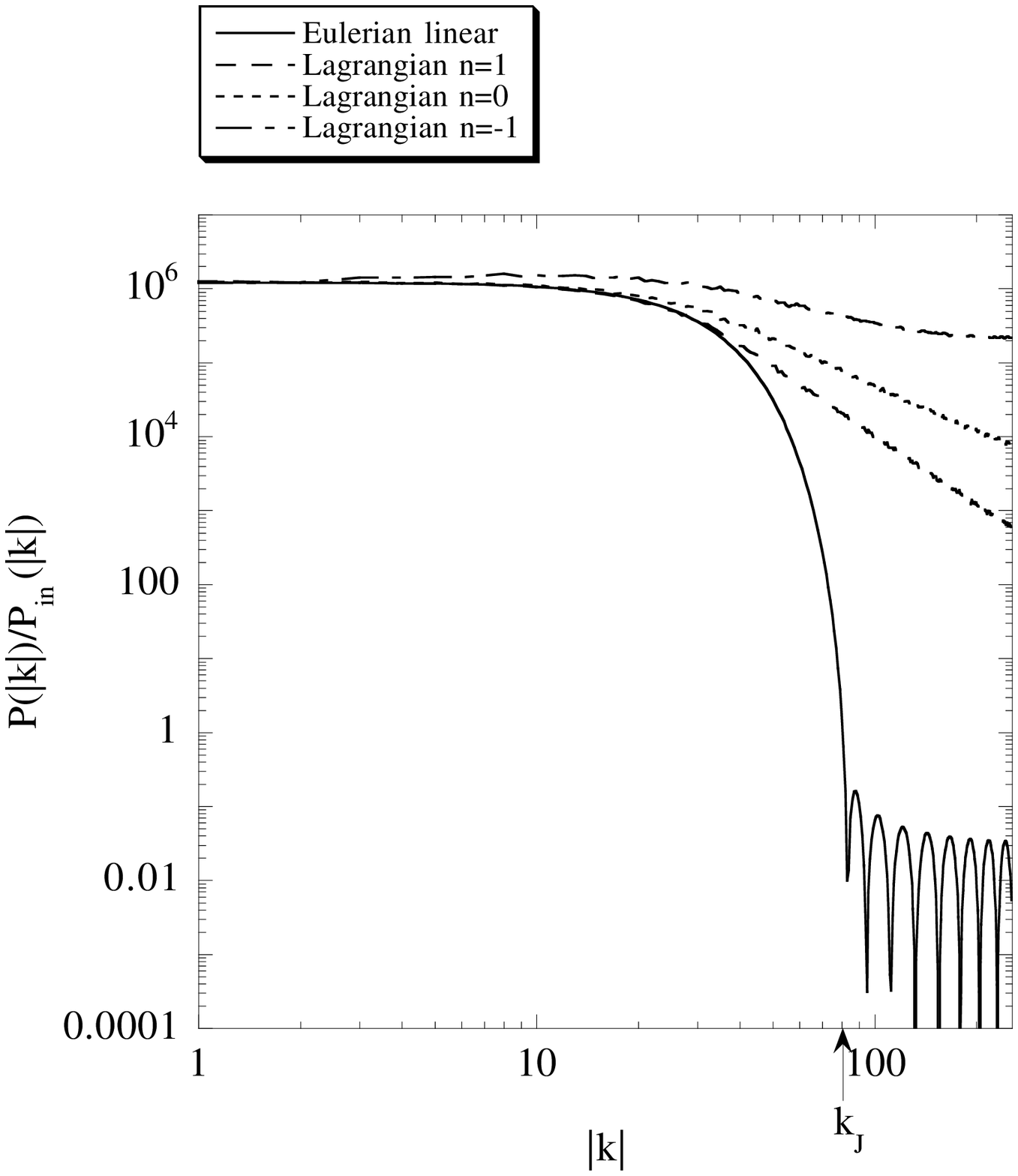}}
\caption{Same as in Fig.~\ref{trans0300}, but for $a=1100$.}
\label{trans1100}
\end{minipage}
\end{figure}

\begin{figure}
\begin{minipage}{.45\linewidth}
\leavevmode
\epsfysize=80mm
\centerline{\epsfbox{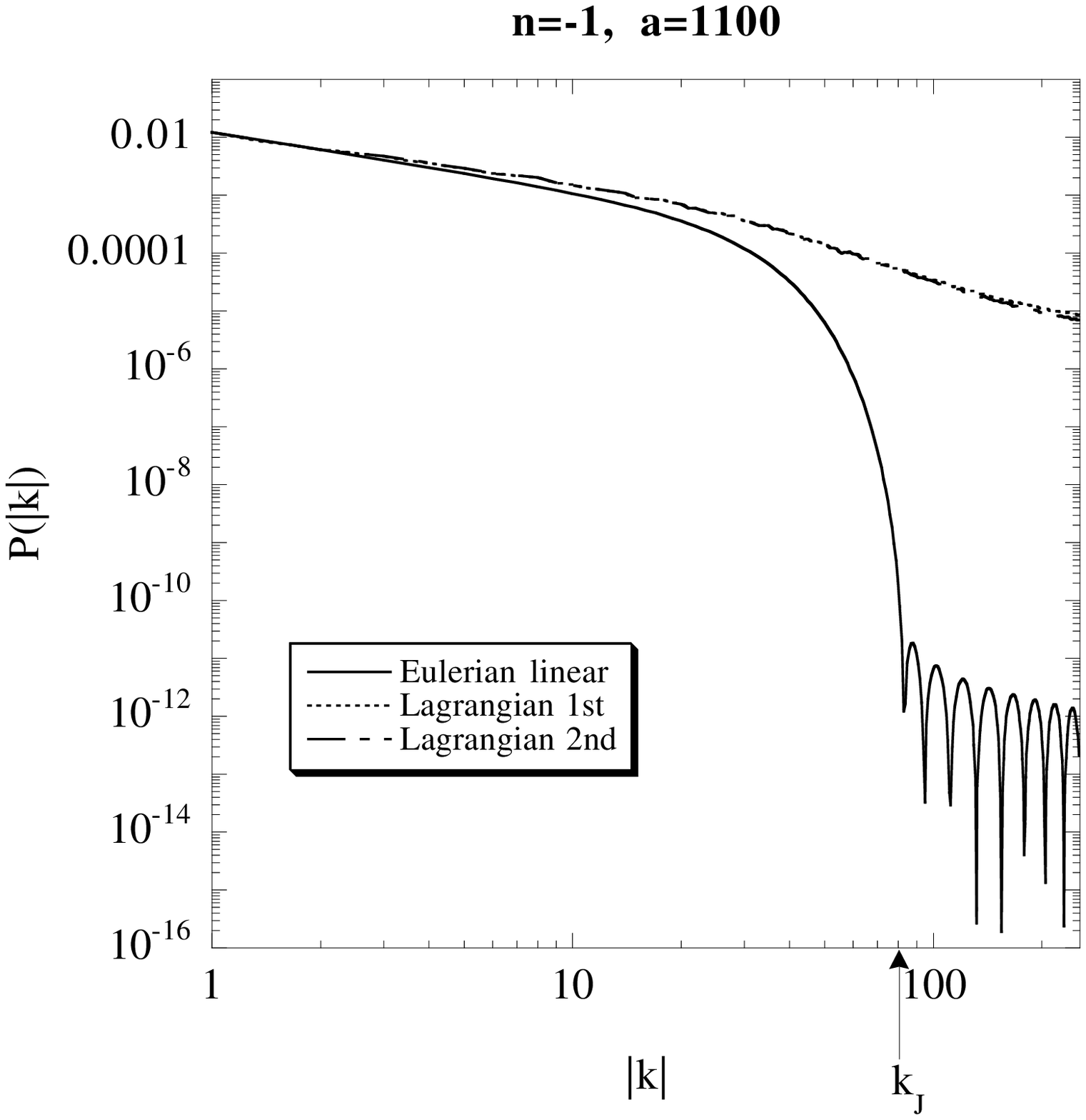}}
\caption{Power spectra of density perturbations at $a=1100$
         computed from the Eulerian linear theory,
         and the Lagrangian first-order and second-order approximations.
         The spectral index is $n=-1$.}
\label{n-1a1100}
\end{minipage}
\begin{minipage}{.05\linewidth}
\quad
\end{minipage}
\begin{minipage}{.45\linewidth}
\leavevmode
\epsfysize=80mm
\centerline{\epsfbox{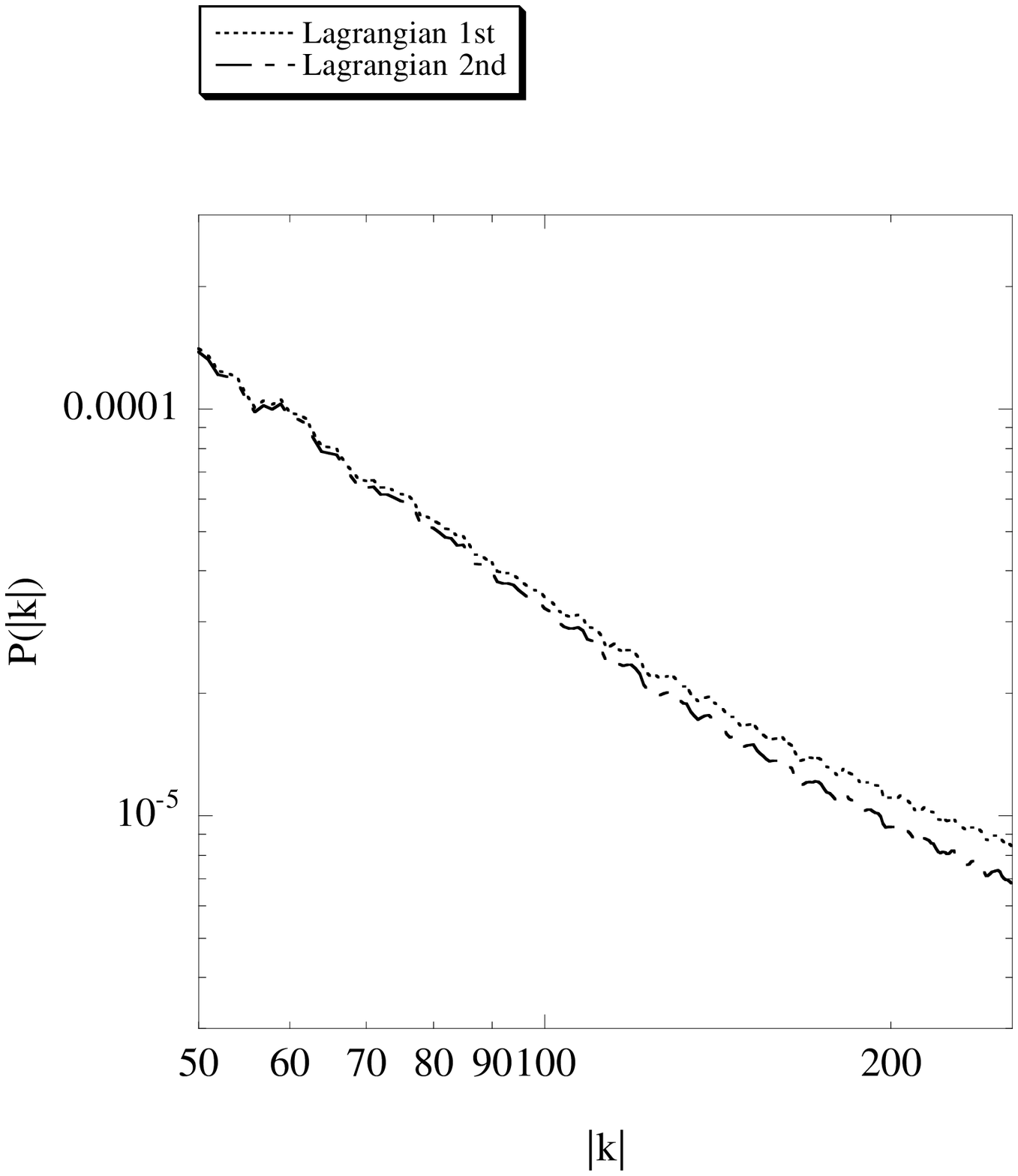}}
\caption{Power spectra at $a=1100$ for $n=-1$,
showing small difference between the results by the first-order and
the second-order Lagrangian approximations.
The spectrum by the Eulerian linear approximation is omitted.}
\label{diff1100}
\end{minipage}
\end{figure}

Now we compute power spectra
${\cal P}(|k|,t) = {\cal P}(k,t) \equiv \langle |\delta(k,t)|^2 \rangle$
of density perturbations,
where $\langle \cdot \rangle$ denotes ensemble average
over the entire distribution,
by using the Lagrangian perturbation theory
formulated in the preceding section.
We also compute them by the Eulerian linear theory
and compare the results to clarify difference of
the Eulerian and the Lagrangian perturbative approximations.
In the Lagrangian approximations, we need some computation
to obtain the power spectra
although the Eulerian approximation yields them directly.
The procedure of the computation is the following:
\begin{enumerate}
\item  First we specify initial conditions as we mentioned above.
       Then we have complete form of the perturbation solutions in $K$-space.
\item  Next we transform the perturbation solutions in $K$-space
       into those in $q$-space via the inverse Fourier transformation.
\item  From the perturbation solutions in $q$-space,
       we immediately find density perturbations in $q$-space
       by equation~(\ref{rho1dim}).
\item  We evaluate density perturbations in $x$-space from those in $q$-space.
\item  Finally by the Fourier transformation with respect to $x$,
       we obtain power spectra of density perturbations.
\end{enumerate}
By way of this procedure, we obtain the power spectra
of density perturbations
presented in Figs~\ref{trans0300}--\ref{diff1100}.
We choose a spectral index as
$n=+1$, 
$0$, 
and $-1$. 
The constant $B$, which is proportional to $\kappa$ and then
provides strength of pressure effect, is put by hand.
Here it is chosen so that the Jeans wavenumber $k_{\rm J} = \sqrt{2/(3B)}$
is $80$.
Note that $k_{\rm J}$ is now a constant because of our choice
of the polytropic index $\gamma=4/3$,
although $k_{\rm J}$ depends on time in general.
We set initial conditions at $a=1$, and pursue the evolution
up to $a=1100$.

Indeed, for all the cases $n=+1$, $0$, and $-1$,
shell crossing will occur at $a \sim 1100$
in the Lagrangian approximations
despite the presence of pressure effect.
(In other words, we normalize amplitude of initial density fluctuations
so that shell crossing occurs at $a \sim 1100$
in the Lagrangian approximations.)
Then we cannot follow the evolution so deeply
into non-linear regime,
as long as we consider the stages before the occurrence
of shell crossing.
Actually in the illustration,
${\cal P}(|k|) \sim 10^{-4}$--$10^{-5}$ at $a=300$,
and ${\cal P}(|k|) \sim 10^{-3}$--$10^{-4}$ at $a=1100$.
Hence we must say that density perturbations remain
in weakly non-linear regime (or in linear regime),
rather than in non-linear regime, through this illustration.

Figs~\ref{trans0300} and \ref{trans1100} show our results
obtained by the Eulerian linear theory and
the Lagrangian first-order approximation,
in terms of the `transfer function,'
${\cal P}(|k|,t) / {\cal P}_{\rm in}(|k|)$.
It is convenient to use it
because it does not depend on the initial conditions
in the Eulerian linear theory, which actually yields
\begin{equation}
\frac{{\cal P}_{\rm lin}(|k|,t)}{{\cal P}_{\rm in}(|k|)}
= \frac{1}{4} \left| \left(1+\frac{5}{6b(k)}\right) D^+(k,t) +
                     \left(1-\frac{5}{6b(k)}\right) D^-(k,t)
              \right|^2 \,.
\end{equation}
In the Lagrangian approximations, however,
it generally depends on the initial conditions.
We do not present in Figs~\ref{trans0300} and \ref{trans1100}
the results by the Lagrangian second-order approximation,
because they are almost coincident with those by the first-order one.
To show the difference between the Lagrangian first-order and
second-order approximations,
we show in Figs~\ref{n-1a1100} and \ref{diff1100}
the power spectra at $a=1100$ for $n=-1$,
where the difference is largest within our calculations.
First we observe the results by the Eulerian linear theory.
The power spectra presented in Figs~\ref{trans0300} and \ref{trans1100}
show the very behaviour of linear density perturbations
stated in subsection 4.1,
i.e. density perturbations with wavenumbers $|k|<k_{\rm J}$ grow
while those with wavenumbers $|k|>k_{\rm J}$ decay with acoustic oscillation.
On the other hand, in the Lagrangian approximations,
the shape of the curves is manifestly different from that
in the Eulerian linear theory.
Although there is little difference on large scales ($|k|<k_{\rm J}$),
we see that amplitude on small scales ($|k|>k_{\rm J}$)
in the Lagrangian approximations is larger than that
in the Eulerian linear theory.
This fact has been observed also in the pressureless case \cite{scba}.
The difference is small at $a=300$, but becomes larger as time proceeds.
Comparing the Lagrangian first-order and second-order approximations
in Fig.~\ref{n-1a1100},
they are found to give almost coincident results through our computation
from $a=1$ to $a=1100$ and
we can hardly observe difference between them.
We present the enlarged power spectra in Fig.~\ref{diff1100},
where the difference is barely visible.
Indeed the difference is less than $10\%$ at $|k| \la 150$.
In the second-order approximation, however,
amplitude of the power spectrum is slightly suppressed,
compared with the first-order one.
The features of the power spectra mentioned above are common
for all the cases, $n=+1$, $0$, and $-1$.

\subsection{Discussions on the power spectra}

Let us consider the reasons of the features of the power spectra.
First we examine the difference between the Eulerian and the Lagrangian
approximations,
which has been discussed in the pressureless case
by Schneider \& Bartelmann \shortcite{scba}.
In the Eulerian linear theory, density perturbations with a wave mode
evolve without being influenced by those with another mode.
In the Lagrangian approximations, however, non-linearity are induced
in calculation of density perturbations.
Origin of non-linearity exists in an expression of the density contrast
in the Lagrangian description as well as in the transformation of
the density contrast in the Lagrangian coordinates $\bmath{q}$
into those in the Eulerian coordinates $\bmath{x}$.
To see this fact, let us express the Lagrangian density perturbations
in the Eulerian coordinates in a one-dimensional system
within a perturbative manner.
The relation between $\bmath{x}$ and $\bmath{q}$ is given as
\[
x_1 = q_1 + s_1(q_1,t) \,,
\]
where we assume that $s_1$ is small enough to be treated as a perturbation.
Using this relation iteratively, the inverse relation is obtained as
\[
q_1 = x_1 - s_1(q_1,t) = x_1 - s_1(x_1,t) + O \left((s_1)^2\right) \,.
\]
Then we have
\begin{equation}\label{s1q-s1x}
s_1(q_1,t) = s_1(x_1,t)
           - \frac{\partial s_1(x_1,t)}{\partial x_1} s_1(x_1,t)
           + O \left((s_1)^3\right)\,,
\end{equation}
\begin{equation}\label{partq1-partx1}
\frac{\partial}{\partial q_1} = \frac{\partial x_1}{\partial q_1}
                                \frac{\partial}{\partial x_1}
= \frac{\partial}{\partial x_1}+\frac{\partial s_1(q_1,t)}{\partial q_1}
                                \frac{\partial}{\partial x_1}
= \frac{\partial}{\partial x_1}+\frac{\partial s_1(x_1,t)}{\partial x_1}
                                \frac{\partial}{\partial x_1}
+ \cdots \,.
\end{equation}
By using equations~(\ref{s1q-s1x}) and (\ref{partq1-partx1}),
the expression of the density contrast in the Lagrangian coordinates
\[
\delta_{\rm L}(q_1,t) = \left(1+s_{1,1}(q_1,t)\right)^{-1} -1
                      = -\frac{\partial s_1(q_1,t)}{\partial q_1}
       +\left( \frac{\partial s_1(q_1,t)}{\partial q_1} \right)^2
       + O \left((s_1)^3\right)
\]
(a subscript `L' denotes `Lagrangian')
is transformed as
\begin{equation}\label{deltaL-x}
\delta_{\rm L}(x_1,t) = -\frac{\partial s_1(x_1,t)}{\partial x_1}
         +\frac{\partial^2 s_1(x_1,t)}{\partial x_1^2} s_1(x_1,t)
         +\left(\frac{\partial s_1(x_1,t)}{\partial x_1}\right)^2
         +O \left((s_1)^3\right) \,,
\end{equation}
where the first term of the right-hand side
corresponds to the density contrast $\delta_{\rm lin}(x_1,t)$
in the Eulerian linear theory.
Equation~(\ref{deltaL-x}) indicates that the density contrast
obtained by the Lagrangian description includes extra non-linear terms,
which cause mode couplings.
For example, if initial density perturbations
consist of a single wave mode,
$\delta_{\rm in}(x_1) \propto \sin kx_1$,
such non-linear terms generate high-frequency modes such as $\sin 2kx_1$.
Of course, in the Eulerian linear theory,
mode couplings never occur
and existence of just a single mode is preserved, i.e.
$\delta_{\rm lin}(x_1,t) \propto \sin kx_1$.
To see the mode-coupling effect quantitatively,
one should consider the Fourier transform of equation~(\ref{deltaL-x}),
\begin{equation}\label{deltaL-k}
\delta_{\rm L}(k,t) = \delta_{\rm lin}(k,t) + \sum_{k' \ne k}
\frac{k \delta_{\rm lin}(k',t) \delta_{\rm lin}(k-k',t)}{k-k'}
+ O \left((\delta_{\rm lin})^3\right) \,,
\end{equation}
where the second term of the right side represents the mode couplings.
Note that the second term is written as the summation
with respect to all wavenumbers.
Then the effect of the second term may be larger than the simple square
of $\delta_{\rm lin}(k,t)$ by order of the number of wave modes.
In the case of our illustration, it may be larger by 2 order.
 For example, if the power spectrum has a peak value $10^{-4}$,
the mode-coupling effect can generate the amplitude of $10^{-6}$
at high frequency.
One may wonder, at first glance of our results,
why the results by the Lagrangian approximations
contain so large amplitude at high frequency,
although the illustration is performed in nearly linear regime.
However, this fact is explained by the effect of
the mode couplings.
The appearance of the large amplitude on small scales
may be interpreted physically as follows.
In the Lagrangian description of hydrodynamics,
one obtains physical quantities in a frame comoving with flow lines of fluid.
If there exists growing density enhancement in a region,
one can see that flow lines there become close to each other
because of gravitational instability.
In other words, one knows by following flow lines
that a physical wavelength of inhomogeneity gets small
due to gravitational contraction.
Actually, density perturbations with an initially small wavenumber $|K|$
become those with a large $|k|$ later.
It can easily seen by the relation between $K$ and $k$, given as
\[
|k| \sim \frac{1}{\ell} \sim \frac{1}{L+s_1(L,t)}
       = \frac{1}{L} \left(1+\frac{s_1(L,t)}{L}\right)^{-1}
    \sim |K| \left(1+\delta(L,t)\right) \,,
\]
where $\ell$ is a physical wavelength of inhomogeneity
measured in the Eulerian coordinates,
and $L$ denotes an initial wavelength.
Thus we may conclude that the appearance of the large amplitude
on small scales is due to the fact that scale of inhomogeneity
is shortened as inhomogeneity grows because of gravitational instability.
Furthermore, let us make sure of behaviour of the power spectra
by the Lagrangian approximations for large $|k|$.
As we stated in the previous subsection,
shell-crossing singularities arise in spite of the presence
of pressure effect in our perturbation scheme.
In our illustration, they arise at $a \sim 1100$,
and thus the epoch $a=1100$ is just before the occurrence
of shell crossing.
In such an epoch,
the power spectrum behaves like ${\cal P}(|k|) \propto |k|^{-1}$
 for large $|k|$ in a one-dimensional system \cite{scba}.
This behaviour concerns only the occurrence of shell crossing,
and is seen not only in the Zel'dovich approximation
but also in our results, Figs~\ref{trans1100} and \ref{n-1a1100}.
Note that in Fig.~\ref{trans1100}, the `transfer function' behaves like
${\cal P}(|k|) / {\cal P}_{\rm in}(|k|) \sim |k|^{-(n+1)}$
at high frequency, showing dependence on initial conditions,
while Fig.~\ref{n-1a1100} shows the behaviour directly.
It should be also stressed that the above three kinds of the arguments
on the Lagrangian power spectra hold true,
whether the pressure effect is taken into account or not.
In this sense, it is natural that the results
by Schneider \& Bartelmann \shortcite{scba} and ours
have similar tendency.
Next let us confirm little difference between the first-order
and the second-order Lagrangian approximations.
 For a rough estimation,
we consider perturbations with a single wave mode $K$
so that the first-order solution in the $q$-space is written in the form
\begin{equation}\label{singlewaveS}
S(q_1,t) = \frac{\epsilon}{K^2} D(K,t) \sin Kq_1 \,,
\end{equation}
where $\epsilon$ denotes amplitude of initial density perturbations,
and $D(K,t)=t^{-1/6+b(K)}$.
Then the second-order solution becomes
\begin{equation}
\zeta(q_1,t) \sim -\frac{\epsilon^2}{4\pi K_{\rm J}^2}
                  D(K,t)^2 \sin 2Kq_1 \,,
\end{equation}
where $K_{\rm J}=k_{\rm J}=\sqrt{2/(3B)}$ is a wavenumber
corresponding to the Jeans length.
The fraction of $S(q_1,t)$ and $\zeta(q_1,t)$ is estimated as
\begin{equation}\label{fraczetaS}
\left| \frac{\zeta(q_1,t)}{S(q_1,t)} \right|
\la \epsilon
\left(\frac{K}{K_{\rm J}}\right)^2 D(K,t) \,.
\end{equation}
If $K>K_{\rm J}$, the factor $D(K,t)$ decreases as time and then
the fraction $|\zeta/S|$ remains small forever.
In contrast, if $K \ll K_{\rm J}$,
the factor $D(K,t)$ increases as time, but $K/K_{\rm J}$ is small.
Then the fraction $|\zeta/S|$ cannot grow to be so large in early time.
Indeed in our calculations, it is less than about $10\%$
during a period up to $a=1100$.
As time proceeds, however, it will become large if $K \ll K_{\rm J}$.
This is a simple argument, but inequality~(\ref{fraczetaS}) may be useful
to give a criterion of effect of the second-order terms.
To let this argument more rigorous,
we should take into account the mode-coupling effect,
as we do in equation~(\ref{deltaL-k}).
In the estimation of the second-order solution,
however, it is not essential to include the mode-coupling effect.
It is rather significant to notice that the right side of
inequality~(\ref{fraczetaS}) has the factor $(K/K_{\rm J})^2$.
The presence of this factor is due to the fact
that the second-order solution is of purely pressure origin
in a one-dimensional model.
In other words, gravitational effect is completely included
in the first-order solution in a one-dimensional model.
Thus we can confirm small difference between the first-order
and the second-order Lagrangian approximations.
This fact tells us that difference between the first-order approximation
and an exact solution is also small, at least up to $a=1100$.
On the other hand, in the pressureless case,
the first-order approximation (i.e. the Zel'dovich approximation)
becomes an exact solution in a one-dimensional model,
and this fact is a strong ground of the validity
of the Zel'dovich approximation.
In this sense, we can also expect accurate description
by the Lagrangian approximations in weakly non-linear regime
in the presence of pressure, as in the pressureless case.

\section{Concluding remarks}

We have developed a perturbative approximation theory,
based on the Lagrangian description of hydrodynamics
in the framework of the Newtonian cosmology,
by extending the method of Adler \& Buchert \shortcite{adler}.
Including `pressure' effect of fluid,
we have derived and solved perturbation equations
in the Lagrangian coordinates up to second order.
Especially we presented explicit form of the second-order
solutions for the case $P \propto \rho^{4/3}$.
We have also computed the evolution of the power spectra
of density perturbations in a one-dimensional model,
based on the Eulerian and the Lagrangian approximations.
Comparing the power spectra,
we have found difference of these approximations.
In particular, large amplitude on small scales has appeared
in the results of the Lagrangian approximations beyond linear regime.
Moreover, the first-order and the second-order Lagrangian approximations
have been found to yield almost the same results within our calculations.
Then we can conclude that, in a one-dimensional system,
the first-order Lagrangian approximation provides
nearly exact description in weakly non-linear regime.

In the computation of the power spectra of density perturbations
by the Lagrangian approximations,
we have found that the shell-crossing singularities occur
even in the presence of the pressure effect.
However, the epoch of the occurrence of shell crossing
in our approximations is, of course,
later than that in the Zel'dovich approximation.
This fact is easily seen by considering perturbations
with a single wave mode so that the first-order solution is given
by equation~(\ref{singlewaveS}) again.
Then the energy density, equation~(\ref{rho1dim}), becomes
\begin{equation}
\rho (q_1,t) = \frac{\rho_{\rm b}(t)}{1-\epsilon D(K,t) \sin Kq_1} \,.
\end{equation}
If $K<K_{\rm J}$, the denominator of the right side goes to zero
in a finite time, i.e. shell crossing will occur.
But, since the growth rate $D(K,t)$ is weaker
than that of the Zel'dovich approximation,
the shell-crossing epoch becomes later.

In this paper, we focus our attention on the case
$P \propto \rho^{4/3}$,
where the solutions of the perturbation equations are written
in a simple form.
This equation of state is nothing but an assumption
to simplify the perturbation solutions.
However, it is crucial what form an effective equation of state
takes when velocity dispersion is replaced with pressure-like force.
Thus, in order to let our formulation more useful,
we must reconsider an equation of state which holds
effectively in high-density regions,
where velocity dispersion plays an important role.
 For example, Buchert \& Dom\'{\i}nguez \shortcite{budo} found
that a relation $P \propto \rho^{5/3}$ is favoured for
small velocity dispersion
under the kinematical restriction that the fluid motion
involves no shear.
Extensions of our formulation to such cases,
as well as more generic determination of an equation of state,
will be the subjects of future investigation.

\section*{Acknowledgments}

We would like to thank the referee, Professor Bernard Jones,
for constructive comments.
We also thank Thomas Buchert, Kei-ichi Maeda, Hiroki Anzai,
and Momoko Suda for helpful discussion and many valuable remarks.

\label{lastpage}
\end{document}